\documentstyle[12pt,epsfig]{article}
\begin{document}
\begin{titlepage}
\begin{center}
{\large \bf An alternative way of plotting the data and 
results of models of $J/\psi$ suppression}
\end{center}
\vspace{1cm}
\begin{center}
{ J\'an Pi\v{s}\'ut${}^{a)}$ and Neva Pi\v{s}\'utov\'a${}^{b)}$}
\end{center}
\vspace{1cm}
\begin{center}
${}^{a)}${\it Theory Division, CERN, CH-1211 Geneva 23, Switzerland,

 on leave of absence from ${}^{b)}$}
\end{center}
\begin{center}
${}^{b)}${\it Department of Physics, Comenius University,

 SK-84248 Bratislava, Slovakia}
\end{center}

\vspace{1cm}
\abstract
{ We propose an alternative way of looking at data on 
anomalous $J/\psi$ suppression. The proposed method is 
in principle equivalent to the one used by the NA50 Collaboration,
but it permits to visualize separate contributions
 of individual processes
  responsible  for the disintegration  
 of $J/\psi$'s  produced by a hard process in nuclear collisions.
The method can be used provided that the time sequence 
of contributing
mechanisms is known or assumed. It offers 
an alternative graphical presentation of the 
onset of anomalous $J/\psi$ suppression in Pb--Pb 
interactions observed
by the NA50 Collaboration at the CERN SPS and might contribute to
explain why different mechanisms, such as $J/\psi$ suppression by the 
Quark--Gluon Plasma and by co--movers in the Dual Parton
Model or in Monte Carlo microscopic approaches, all lead to an
approximate description of anomalous $J/\psi$ suppression.}
\end{titlepage}
\section{Introduction}
\label{intro}

The anomalous $J/\psi$ suppression observed
by the NA50 Collaboration \cite{NA50a,NA50b} 
in Pb--Pb collisions at the CERN SPS can be naturally 
interpreted as $J/\psi$ dissolution 
by the Quark--Gluon Plasma (QGP). This phenomenon was  
predicted by Matsui and Satz \cite{MS86} more than 15 years ago.
The interpretation of the anomalous $J/\psi$ suppression is 
somewhat complicated by the fact that there are also other 
contributions to $J/\psi$ suppression. Nuclear absorption (NA) 
(sometimes referred to as ''pre-resonance absorption'' or 
''Gerschel--H\"ufner mechanism'')
\cite {GH88}--\cite{GH99} is probably responsible for most of $J/\psi$ 
suppresion in
nuclear collisions induced by lighter ions. Disintegration
 of $J/\psi$ by collisions with secondary hadrons 
\cite {BM}--\cite{Kaid} 
(also referred to as ''co-mover'' interaction) 
is another possible contribution. In order to make the 
$J/\psi$ suppression by the gas of secondary 
hadrons sufficiently large,
  in some approaches \cite{Spiel,Vogt2}
cross-section ($\sigma_{co}$) for the process 
hadron+$J/\psi \to D\bar D+X$ has to be of about $5$ mb, depending
also on the value of the cross-section ($\sigma_a$) for nuclear
absorption. In the approach based on the Dual Parton Model (DPM)
\cite{Arm} one needs $\sigma_{co}=0.6$ mb, for $\sigma_a=6.7$ mb and
\cite{Kaid} $\sigma_{co}=1$ mb for $\sigma_a=4.5$ mb.
 It has been argued \cite{KShadx} that, at energies corresponding to 
the thermal motion of hadrons, the cross-sections for the 
disintegration of $J/\psi$ by hadrons are
in fact an order of magnitude smaller. The issue is not yet 
definitely clarified.

There is one key difference between the 
suppression of $J/\psi$ by 
the QGP and by other mechanisms. For $J/\psi$ 
dissolution by QGP a  rapid onset of the 
anomalous suppression is expected or at least possible, 
whereas for other mechanisms one
 expects a smooth dependence of $J/\psi$ suppression 
on the impact parameter $b$ and on the total transverse energy $E_T$
 of a nuclear collision. Models by 
Blaizot and Ollitrault (BO) \cite{BO96} and by 
Kharzeev, Louren\c{c}o, Nardi and Satz (KLNS) \cite{KLNS97,satz}
 do show a rather rapid onset of the dependence 
of the anomalous $J/\psi$ suppression on  
 $E_T$ or on $b$. Plots of experimental data 
on $J/\psi$ suppression, in particular those with the 
nuclear absorption ''subtracted'' \cite{NA50c} 
do show such an abrupt onset.

Although the anomalous $J/\psi$ suppression is a rather spectacular
Phenomenon, indicating a rapid onset of a new mechanism of $J/\psi$
Suppression, there are a few mechanisms that are able to obtain 
some agreement with the basic features 
of the data. Apart from the suppression by QGP
\cite{BO96,KLNS97,satz} these include the disintegration of $J/\psi$
by co-movers \cite{Arm,Kaid} and Monte Carlo microscopic models 
\cite{Spiel}. In order to see why it is possible that rather 
different models lead to a similar overall $J/\psi$ suppression, it
would be interesting to see not only the resulting
total $J/\psi$ suppression but also contributions of
individual mechanisms involved in this total suppression.
The purpose of the present paper is to propose an alternative way
of looking at the data on $J/\psi$ suppression, and in particular 
on the onset of anomalous $J/\psi$ suppression. Our approach is
 based on the additive
decomposition of contributions of different
 mechanisms responsible for
the disintegration of $J/\psi$'s produced by a hard process in nuclear
collisions. The method can be used provided that the time sequence 
of contributing mechanisms is known or assumed. In some cases there 
are good reasons to believe that two mechanisms are at work 
in the same time period. This would concern, for instance, the 
nuclear absorption and possible depletion of gluon 
structure functions \cite{Hwa} 
during the first part of the nuclear collision. In such a case both 
mechanisms should be considered as one stage.

In the next section we shall describe the method; in Sect. 3 we shall
present a few illustrative examples. In Sect. 4
we shall discuss  the relationship of the proposed way of 
looking at the data with the standard procedure used by 
the NA50 Collaboration, and comments 
and conclusions will be presented in Sect. 5.

\section {Separation of different contributions
to $J/\psi$ disintegration}
\label{method}     

We shall consider here for simplicity the case when only two 
mechanisms of $J/\psi$ suppression are present: 
the nuclear absorption and the dissolution of $J/\psi$ by 
QGP, the latter taken from a
point of view very close to that of the BO \cite{BO96} and KLNS 
\cite{KLNS97,satz} models.

We shall use the following notation

\medskip
$N^{J/\psi}_{prod}(E_T)=$ the total number of $J/\psi$ produced 
in collisions at given $E_T$,

\medskip
$N^{J/\psi}_{NA}(E_T)=$ the number of $J/\psi$'s disintegrated by NA, 

\medskip
$N^{J/\psi}_{QGP}(E_T)=$ the number of $J/\psi$'s dissolved by QGP,

\medskip

$N^{J/\psi}_{exp}(E_T)=$ the number of surviving $J/\psi$'s,
 measured in experiment 

\medskip
$N^{J/\psi}_{other}(E_T)=$ the numbers of $J/\psi$'s disintegrated
 by other mechanisms.

\medskip
The balance between input and output requires
\begin{equation}
N^{J/\psi}_{prod}(E_T)= N^{J/\psi}_{exp}(E_T)+
 N^{J/\psi}_{NA}(E_T)+ N^{J/\psi}_{QGP}(E_T)+ N^{J/\psi}_{other}(E_T).
\label{eq1}
\end{equation}
We suggest that all terms in Eq. (1) be calculated 
and plotted separately, or 
in combinations, for a given A-B collision, for all values of $E_T$.
In this sense our approach tries to give an answer 
to the question ''Where are all the
$J/\psi$'s gone?'' or corresponds to ''an accountant's look at the data 
on $J/\psi$ suppression''.

Admittedly, the suggestion we are making is rather trivial, but we 
do hope that looking at the data in this way can make the results
of a particular model more transparent and can help to avoid some
inconsistencies.

In specific applications Eq.(1) can be rewritten in such a way that
more reliably known terms are put on one side, 
thus providing  constraints
for the more model-dependent and less reliably known ones. We can for 
instance put on one side the best known terms 
$N^{J/\psi}_{prod}(E_T)- N^{J/\psi}_{exp}(E_T)$ or even
$N^{J/\psi}_{prod}(E_T)- N^{J/\psi}_{exp}(E_T)- N^{J/\psi}_{NA}(E_T)$,
and try to describe or fit these expressions by a particular model.
For instance in the latter case by 
$N^{J/\psi}_{QGP}(E_T)+ N^{J/\psi}_{other}(E_T)$.

The purpose of the present paper is not to attempt
a detailed analysis of the
data. We wish just to describe the method and to give a few 
examples as illustrations. For that purpose we shall take nuclei as 
homogeneous hard spheres with radii $R_A=1.2A^{1/3}$fm, 
we shall neglect the 
energy fluctuations, including the ''knee'' in multiplicity distribution 
at $E_T\approx 100$GeV and we shall assume here a 
strong correlation between $b$ and $E_T$:  
\begin{equation}
E_T(b)=0.325 GeV N_w(b),
\label{eq2}
\end{equation}
where $N_w(b)$ is the number of interacting (''wounded'') nucleons.
With these simplifications the quantities entering Eq. (1) can be 
expressed in the following way:
\begin{equation}
N^{J/\psi}_{prod}(E_T)=\sigma^{J/\psi}_{nn} N_{coll}(E_T),
\label{eq3}
\end{equation}
where $\sigma^{J/\psi}_{nn}$ is the $J/\psi$ production 
cross-section in the average nucleon--nucleon collision and 
$N_{coll}(E_T)$ is the number of nucleon--nucleon collisions at 
a given value of $E_T$. In a rather simplified case we have
$$
N_{coll}(E_T)=
\int_0^{R_A}{\frac{sds}{\sigma_{nn}}}
\int_0^{2\pi}d\theta 
\rho_A\sigma_{nn}2L_A(s)\rho_A\sigma_{nn}2L_B(b,s,\theta)=
$$
\begin{equation}
\int_0^{R_A}{\frac{sds}{\sigma_{nn}}}
\int_0^{2\pi}d\theta 
\int_{-L_A(s)}^{L_A(s)}\rho_A\sigma_{nn}dz_A
\int_{-L_B(s,\theta)}^{L_B(b,s,\theta)}\rho_B\sigma_{nn}dz_B,
\label{eq4}
\end{equation}
where
$$   
L_A(s)=\sqrt{R^2_A-s^2},\quad  
L_B(b,s,\theta)=\sqrt{R_B^2-b^2-s^2+2bscos(\theta)},
$$
when the expression under the square--root in $L_B$ is negative,
there is no tube-on-tube collision and the contribution vanishes. 
Nuclear densities are denoted as $\rho_A,\rho_B$  (we 
are working in the approximation of nuclei as  hard spheres,
therefore $\rho_A,\rho_B$ are constants), $b$ is
the impact parameter, $s$ is the distance from the centre of the 
A-nucleus and $\theta$ is the angle between $\vec b$ and $\vec s$.
A nuclear collision is taken as the sum of tube-on-tube collisions,
with lengths of tubes  $2L_A$ and $2L_B$, and $z_A$,$z_B$ 
specifying the 
coordinate of nucleons within both  colliding tubes: $z_A$ varies
from $-L_A$ to $L_A$, $z_B$ from $-L_B$ to $L_B$, and both $z_A$ and 
$z_B$ increase in the direction of motion of the tubes in the c.m.
frame of the nucleon--nucleon collision.
 The nucleon-nucleon non-diffractive cross-section $\sigma_{nn}$ is
taken as $30$mb.
 Experimental results on the survival probability $S(E_T)$ of $J/\psi$
enter the expression
\begin{equation}
N^{J/\psi}_{exp}(E_T)=S(E_T)N^{J/\psi}_{prod}(E_T).
\label{eq5}
\end{equation}
Nuclear absorption of $J/\psi$ is given by the expression
$$
N^{J/\psi}_{NA}(E_T)=
\sigma^{J/\psi}_{nn}\int_0^{R_A}{\frac{sds}{\sigma_{nn}}}
\int_0^{2\pi}d\theta 
\int_{-L_A(s)}^{L_A(s)}\rho_A\sigma_{nn}dz_A
\int_{-L_B(s,\theta)}^{L_B(b,s,\theta)}\rho_B\sigma_{nn}dz_B
$$
\begin{equation}
(1-e^{-\rho_A\sigma_a[z_A+L_A(s)]}
e^{-\rho_B\sigma_a[z_B+L_B(b,s,\theta)]}),
\label{eq6}
\end{equation}
where $\sigma_a$ is the cross-section describing the absorption 
of $J/\psi$ by nucleons. 

The integral in Eq. (6) can be decomposed into two separate integrals,
the former giving the total number of $J/\psi$'s produced and the 
latter being equal to the number of $J/\psi$'s that survive the
 nuclear absorption.
The term $N^{J/\psi}_{QGP}(E_T)$ gives the number of $J/\psi$'s
destroyed by the QGP; it is given as
$$
N^{J/\psi}_{QGP}(E_T)=
\sigma^{J/\psi}_{nn}\int_0^{R_A}{\frac{sds}{\sigma_{nn}}}
\int_0^{2\pi}d\theta 
\int_{-L_A(s)}^{L_A(s)}\rho_A\sigma_{nn}dz_A
\int_{-L_B(s,\theta)}^{L_B(b,s,\theta)}\rho_B\sigma_{nn}dz_B
$$
\begin{equation}
e^{-\rho_A\sigma_a[z_A+L_A(s)]}
e^{-\rho_B\sigma_a[z_B+L_B(b,s,\theta)]}
\Theta(\kappa-\kappa_{crit}).
\label{eq7}
\end{equation}
Following KLNS \cite{KLNS97}, see also Ref.\cite{NPP}, 
we have introduced 
\begin{equation}
\kappa= \frac{\rho_A\sigma_{nn}2L_A(s).
        \rho_B\sigma_{nn}2L_B(b,s,\theta)}
       {\rho_A\sigma_{nn}2L_A(s)+\rho_B\sigma_{nn}2L_B(b,s,\theta)}.
\label{eq8}
\end{equation}
The parameter $\kappa_{crit}$ specifies the onset of the QGP formation.
 This parameter has to be determined by the data in such a way that 
\begin{equation}
N^{J/\psi}_{prod}(E_T)= N^{J/\psi}_{exp}(E_T)+ N^{J/\psi}_{NA}(E_T)+
N^{J/\psi}_{QGP}(E_T),
\label{eq9}
\end{equation}
where we have assumed that the contribution $N^{J/\psi}_{other}$ 
can be neglected.
Equation (9) simply says that out of all $J/\psi$'s produced 
some survived, others were disintegrated by the nuclear 
absorption and  others by QGP.

The multiplicative factor $\sigma^{J/\psi}_{nn}$ is present in all
of the terms in Eq. (9) and in what follows we shall leave it out. 

The time ordering of the different processes enters Eq. (7).
 It is most likely that the QGP is formed after the colliding nuclei
 have passed through one another  
and the QGP can thus destroy only those $J/\psi$'s that survived 
the nuclear absorption.

\section{Illustrative examples}
\label{data} 

Figures showing the experimental data in the way they were used by NA50 
Collaboration
\cite{NA50a,NA50b} are well known. In these figures they plot 
the survival 
probability of $J/\psi$ as
\begin{equation}
S(E_T)=\frac{J/\psi_{measured}}{J/\psi_{produced}},
\label{eq10}
\end{equation}
inferred from direct data on $J/\psi$ over Drell--Yan pair production. 
The coefficient $S(E_T)$ contains in a multiplicative way
 probabilities that $J/\psi$ has survived nuclear absorption, 
QGP dissolution and other possible mechanisms of $J/\psi$
 disintegration. In the way of looking at data proposed 
here we plot separately the surviving $J/\psi$'s 
(as given by experiment), those
 disintegrated by nuclear absorption, and those 
dissolved by the QGP and possibly 
also by other mechanisms.

\begin{figure}[t]
\vbox{
   \begin {center}
      \epsfig{file=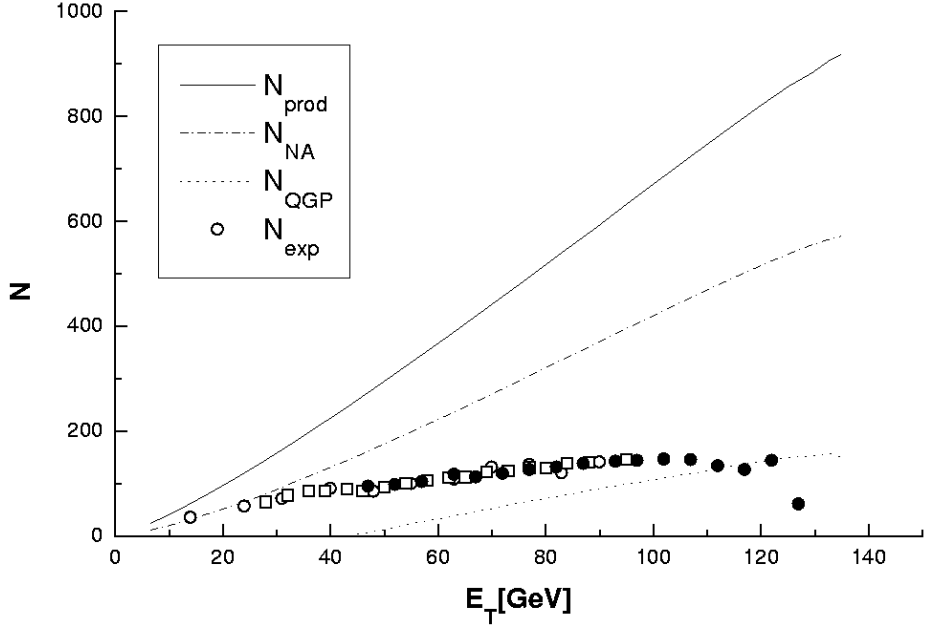,scale=0.5}
   \end {center}
\caption{Results on $N^{J/\psi}_{prod}$ as calculated by Eq.(3)
with $\sigma_{nn}=$3fm${}^2$, the data of NA50 Collaboration 
\cite{NA50a,NA50b} used to calculate  $N^{J/\psi}_{exp}$ 
by Eq. (5) and the  $N^{J/\psi}_{NA}$ calculated by Eq.(6) 
with $\sigma_a=$0.7fm${}^2$, and $N^{J/\psi}_{QGP}$ evaluated
by Eq. (7).
\label{F1}
}
}
\end{figure}

In Fig.~\ref{F1} we plot as an illustration 
the terms $N^{J/\psi}_{prod}$, $N^{J/\psi}_{exp}$ (survivors), and 
$N^{J/\psi}_{NA}$,
leaving out the common factor $\sigma^{J/\psi}_{nn}$ in Eqs. (3)--(7).
The anomaly is not easily visible, since it 
appears as a rapidly growing difference
$N^{J/\psi}_{prod} - N^{J/\psi}_{exp}$.
\begin{figure}[t]
\vbox{
   \begin {center}
      \epsfig{file=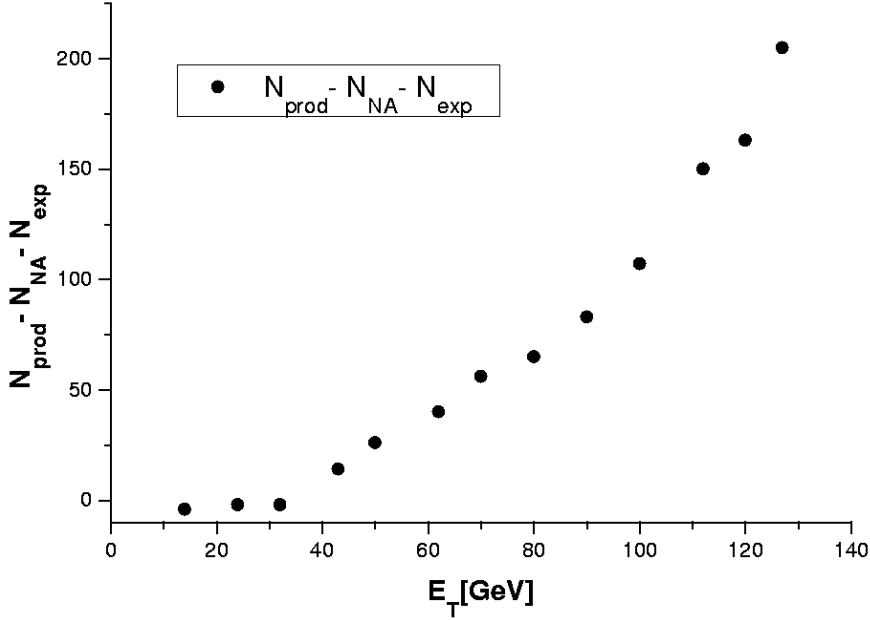,scale=0.5}
   \end {center}
\caption{The expression 
$\Delta N$ = $N^{J/\psi}_{prod}-N^{J/\psi}_{exp}- N^{J/\psi}_{NA}$ 
as a function of $E_T$. The term $N^{J/\psi}_{NA}$ has been 
calculated with 
$\sigma_a=$0.7fm${}^2$.
\label{F2}
}
}
\end{figure}
In order to make the anomaly visible, we plot in Fig.~\ref{F2} the expression
$\Delta N$ = $N^{J/\psi}_{prod}-N^{J/\psi}_{exp}- N^{J/\psi}_{NA}$ 
with a nuclear absorption 
cross-section $\sigma_a=0.7 fm^2$. The shape of $\Delta N$ 
indicates a presence 
of a $J/\psi$ dissolving mechanism with threshold between
 $E_T=30$GeV and 
$E_T=40$GeV. Note that values of $\Delta N$ are slightly 
negative for 
the  lowest three $E_T$ points, which is connected with 
the fact that in the standard 
presentation of data calculations based on the
 NA mechanism are below the lowest
$E_T$ data.

\begin{figure}[t]
\vbox{
   \begin {center}
      \epsfig{file=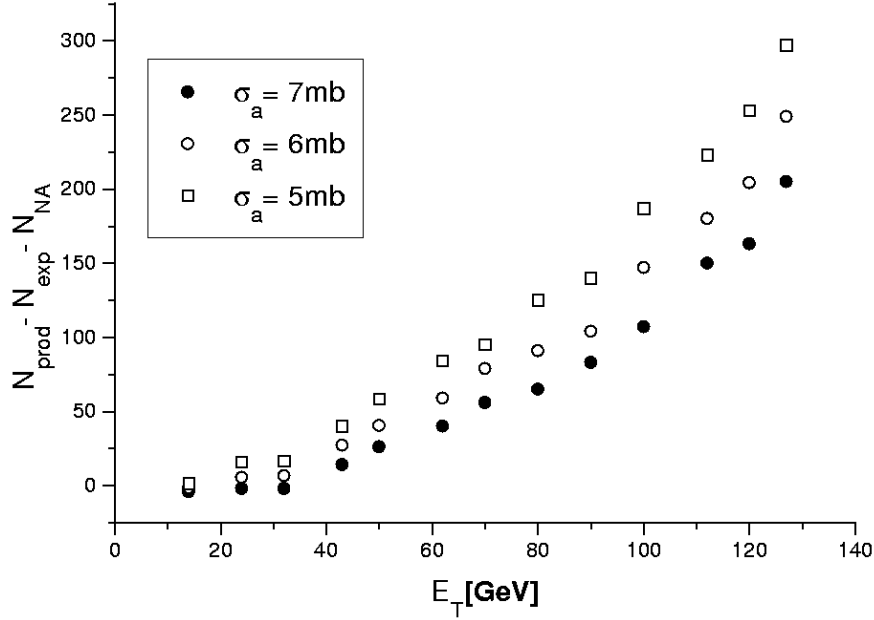,scale=0.5}
   \end {center}
\caption{The expression 
$\Delta N$ = $N^{J/\psi}_{prod}-N^{J/\psi}_{exp}- N^{J/\psi}_{NA}$ 
as a function of $E_T$. The term $N^{J/\psi}_{NA}$ has been 
calculated with $\sigma_a=$0.7fm${}^2$, $\sigma_a=$0.6fm${}^2$, 
and $\sigma_a=$0.5fm${}^2$.
\label{F3}
}
}
\end{figure}

In Fig.~\ref{F3} we show the opening of the space for 
other contributions, most probably for $J/\psi$ 
suppression by a hadron gas, when $\sigma_a$ 
in nuclear absorption becomes smaller.

The true consistency check of the phenomenological description 
of data is provided by Eq. (1) with $N^{J/\psi}_{exp}$ given by
Eq. (5) and $S(E_T)$ taken from experimental
 data on $J/\psi$ survival.
In the illustrative case discussed above we have 
$N^{J/\psi}_{other}=0$ and Eq. (1) reduces to
\begin{equation}
N^{J/\psi}_{prod}(E_T)=\Sigma(E_T)\equiv N^{J/\psi}_{exp}(E_T)+
 N^{J/\psi}_{NA}(E_T)+ N^{J/\psi}_{QGP}(E_T).
\label{eq11}
\end{equation}

\begin{figure}[t]
\vbox{
   \begin {center}
      \epsfig{file=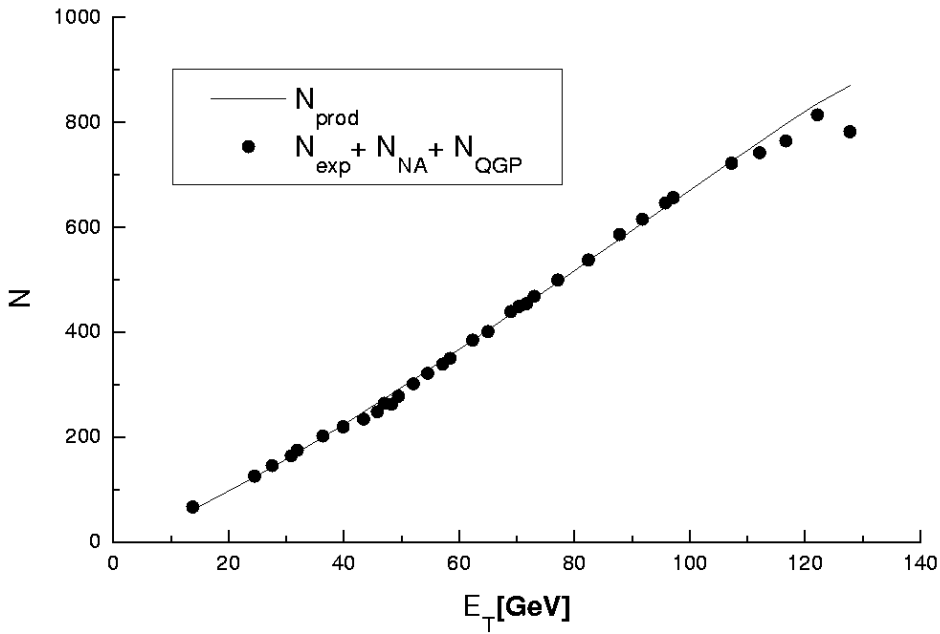,scale=0.5}
   \end {center}
\caption{
The comparison of $N^{J/\psi}_{prod}$ (solid line) and $\Sigma(E_T)$
(open circles).
\label{F4}
}
}
\end{figure}

In Fig.~\ref{F4} we compare $N^{J/\psi}_{prod}(E_T)$ and $\Sigma(E_T)$. The
agreement is quite reasonable, in view of simplifying
assumptions that we made. Agreement with data is worse for the points with the highest 
values of $E_T$ where a large part of $E_T$ is given by fluctuations
\cite{Kaid} and our simplified model is not applicable. Note that
the agreement visible in Fig.~\ref{F4} is non-trivial since 
 $N^{J/\psi}_{exp}(E_T)=S(E_T)N^{J/\psi}_{prod}$ is given
 by experimental data on $S(E_T)$.

The information contained in Fig.~\ref{F1} and Fig.~\ref{F4} 
can be compressed into
\begin{figure}[t]
\vbox{
   \begin {center}
      \epsfig{file=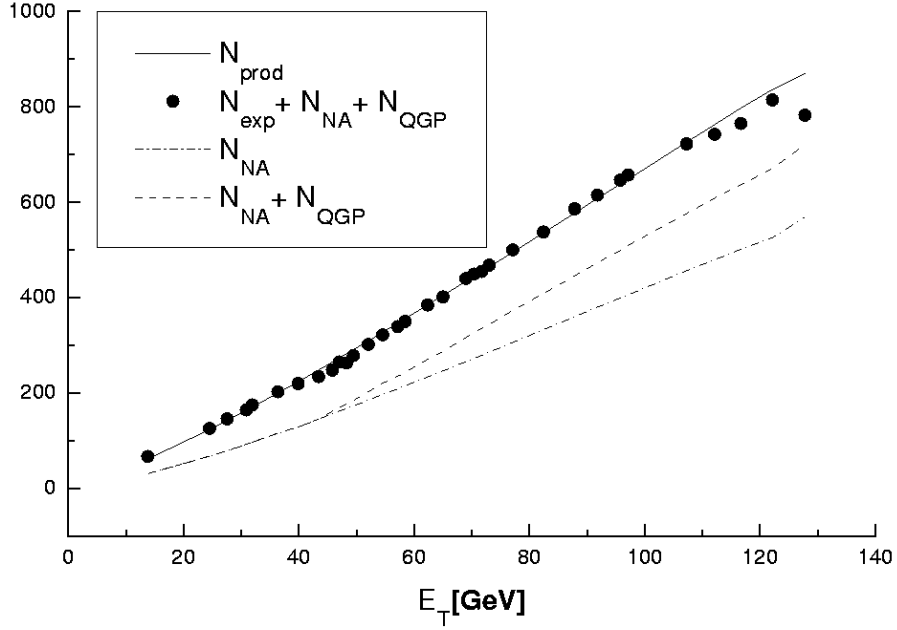,scale=0.5}
   \end {center}
\caption{The compressed information from Figs.~\ref{F1} and \ref{F4}. 
$N^{J/\psi}_{NA}(E_T)$ (dash-dotted line); 
$N^{J/\psi}_{NA}(E_T)+N^{J/\psi}_{QGP}(E_T)$ (dashed line);
$N^{J/\psi}_{NA}(E_T)$+$N^{J/\psi}_{QGP}(E_T)$+$N^{J/\psi}_{exp}(E_T)$
(open circles) and $N^{J/\psi}_{prod}(E_T)$ (solid line).
\label{F5}
}
}
\end{figure}
a single Fig.~\ref{F5}, where we plot $N^{J/\psi}_{NA}(E_T)$;
 $N^{J/\psi}_{NA}(E_T)$+ $N^{J/\psi}_{QGP}(E_T)$ and 
 $N^{J/\psi}_{NA}(E_T)$+ $N^{J/\psi}_{QGP}(E_T)$+
 $N^{J/\psi}_{exp}(E_T)$;
and for comparison with the last expression also 
$N^{J/\psi}_{prod}(E_T)$.

\section{Common points of the new way of looking at data and the
standard procedure}
\label{equivalency}

We shall first show that the way of plotting experimental data 
and results of 
phenomenological calculations is in some aspects equivalent 
to the standard way introduced and used by the NA50 Collaboration.

To make the argument transparent, suppose that a certain number $N_0$
of $J/\psi$'s is passing successively through three 
obstacles (media) and can be disintegrated by each of them.
The probability that it passes (survives) through the first obstacle
is denoted as $P_1$. The number of $J/\psi$'s passing through is 
$N_0P_1$, the number of those that fail to pass is 
$N_{F1}=N_0(1-P_1)$. In the same way the number of $J/\psi$'s passing
through the second obstacle is $N_0P_1P_2$ and the number of those 
that passed the first obstacle, but failed to pass the second one is
$N_{F2}=N_0P_1(1-P_2)$. Finally, the number of $J/\psi$'s passing 
throught the third obstacle is $N_0P_1P_2P_3$ and the number of those
that failed to pass the third obstacle is $N_{F3}=N_0P_1P_2(1-P_3)$.
The obvious identity
\begin{equation}
N_0=N_0(1-P_1)+N_0P_1(1-P_2)+N_0P_1P_2(1-P_3)+P_1P_2P_3N_0
\label{eq12}
\end{equation}
can be rewritten as
\begin{equation}
N_0=N_{F1}+N_{F2}+N_{F3}+SN_0.
\label{eq13}
\end{equation}
where
\begin{equation}
S=P_1P_2P_3=\frac{N_0-N_{F1}-N_{F_2}-N_{F3}}{N_0}
\label{eq14}
\end{equation}
gives the overall survival probability. 
When comparing the results of the calculations with
the data in the usual way 
the comparison is between experimentally measured values of S and the
probability $P_1P_2P_3$ calculated from phenomenological models.
The alternative way used here consists in proceeding 
according to Eq. (12), plotting individual terms 
$N_{F1},N_{F2},N_{F3},SN_0$, with $S$ taken from experiment and 
verifying the validity of Eq.(12).

In a real situation, each $J/\psi$ is born at a certain value 
of the parameters $s,\theta$ (when considering nuclear collisions
at a fixed value of $b$, and at a certain value of $b,s,\theta$
when integrating also over $b$. The probability of passing though
different media depends on values of these parameters. 
 Equations (11--13) change only a little. Lumping $s,\theta$,
or $b,s,\theta$ into a single parameter $x$ and introducing
probabilities $P_1(x)$, $P_2(x)$ and $P_3(x)$, we have in an 
obvious notation:
$$N_0=\int n_0(x)dx$$
$$
N_{P1}=\int n_0(x)P_1(x)dx, \qquad N_{F1}=\int n_0(x)[1-P_1(x)]dx       
$$
$$  
N_{P2}=\int n_0(x)P_1(x)P_2(x)dx, \qquad
 N_{F2}=\int n_0(x)P_1(x)[1-P_2(x)]dx
$$
$$       
N_{P3}=\int n_0(x)P_1(x)P_2(x)P_3(x)dx, \quad
 N_{F3}=\int n_0(x)P_1(x)P_2(x)[1-P_3(x)]dx
$$
and one can write again Eqs. (11)--(13). 

When calculating the total $J/\psi$ suppression via 
an expression corresponding to Eq. (14),
we get:
\begin{equation}
S(E_T)=\frac{N_{prod}^{J/\psi}-N_{GH}^{J/\psi}-N_{QGP}^{J/\psi}}
{N_{prod}^{J/\psi}}.
\label{eq15}
\end{equation}
 Using Eqs. (4)--(6) and the identity 
$1-\Theta(\kappa-\kappa_{crit})=\Theta(\kappa_{crit}-\kappa)$,
we obtain from Eq. (15) the standard expression valid for 
$J/\psi$ suppression by nuclear absorption and by the QGP
$$
S^{J/\psi}(E_T)=
\int_0^{R_A}{\frac{sds}{\sigma_{nn}}}
\int_0^{2\pi}d\theta 
\int_{-L_A(s)}^{L_A(s)}\rho_A\sigma_{nn}dz_A
\int_{-L_B(s,\theta)}^{L_B(b,s,\theta)}\rho_B\sigma_{nn}dz_B
$$
\begin{equation}
e^{-\rho_A\sigma_a[z_A+L_A(s)]}
e^{-\rho_B\sigma_a[z_B+L_B(b,s,\theta)]}\Theta(\kappa_{crit}-\kappa).
\label{eq16}
\end{equation}

\section{Comments and conclusions}
\label{conclusions}

We have desribed here an alternative way of plotting the 
data and model
calculations on $J/\psi$ suppression and presented a few illustrative
examples. A more detailed analysis of $J/\psi$ suppression 
along this path would certainly require using more
 realistic Woods--Saxon nuclear densities, a more 
realistic relationship between $b$ and $E_T$, 
detailed analysis of $J/\psi$ suppression in collisions 
induced by lighter ions at $200$GeV per nucleon and a 
detailed evaluation of experimental errors. 
We are of the opinion that such an undertaking, 
although rather tedious, might help us to understand why models based
on rather different assumptions lead to roughly similar results.

For models considering only nuclear absorption 
and $J/\psi$ interaction with co-movers, as in  Refs. 
\cite{Vogt1,Arm,Vogt2,Kaid} the present scheme requires only minimal
modifications. Equation (9) should be rewritten as
\begin{equation}
N^{J/\psi}_{prod}(E_T)=N^{J/\psi}_{exp}(E_T)+N^{J/\psi}_{NA}(E_T)+
N^{J/\psi}_{co}(E_T).
\label{eq17}
\end{equation}
where the last term corresponds to the number of $J/\psi$'s suppressed
by the interaction with co-movers 
(with the gas of secondary hadrons). The term 
$N^{J/\psi}_{prod}(E_T)$ is given by Eqs. (3) and (4), 
$N^{J/\psi}_{exp}(E_T)$ by Eq.(5), $N^{J/\psi}_{NA}(E_T)$ 
by Eq. (6) and $N^{J/\psi}_{co}(E_T)$ is calculated, in analogy 
to Eq. (7), as
$$
N^{J/\psi}_{co}(E_T)=
\sigma^{J/\psi}_{nn}\int_0^{R_A}{\frac{sds}{\sigma_{nn}}}
\int_0^{2\pi}d\theta 
\int_{-L_A(s)}^{L_A(s)}\rho_A\sigma_{nn}dz_A
\int_{-L_B(s,\theta)}^{L_B(b,s,\theta)}\rho_B\sigma_{nn}dz_B
$$
\begin{equation}
e^{-\rho_A\sigma_a[z_A+L_A(s)]}
e^{-\rho_B\sigma_a[z_B+L_B(b,s,\theta)]}
\Big\lgroup 1-\exp(-\int d\tau
 \langle v\sigma_{co}\rho_{co}\rangle) \Big\rgroup.
\label{eq18}
\end{equation}
In Eq.(18), $\sigma_{co}$ is the cross-section for the disintegration
of $J/\psi$ by the interaction with co-movers,
 $\rho_{co}(\tau)$ is the 
density of co-movers as a function of the proper
 time $\tau$ of $J/\psi$.
For more details see Refs. \cite{FLP,Vogt1,Arm,Vogt2,Kaid,Vogt3}.

For Monte Carlo microscopic models such as the one in 
 Ref.\cite{Spiel}, an analytic
recipe for separating different contributions cannot be written down,
but the separation is most likely possible as well.

The alternative way of plotting and looking at data might be useful,
but it will not resolve the physics problems. The question of 
whether the QGP is responsible for anomalous $J/\psi$ suppression
observed by the NA50 Collaboration will probably and hopefully be
resolved in one of the following ways (or by combination of some of
them):
\begin{itemize}
\item Accurate determination of $\sigma_a$ by data on $J/\psi$ 
suppression at $E_{Lab}=160$GeV per nucleon. Using this value of
$\sigma_a$ together with data on $S(E_T)$ in Pb--Pb interactions
might reveal a threshold of ''missing contribution'',
 when plotted as in Fig.~\ref{F2} above. 
\item Combining data on anomalous $J/\psi$ suppression with 
information on other signatures, including the data \cite{NA50d}
on transverse momenta of surviving $J/\psi$'s and their analyses
\cite{KNS97,HZ1,HZ2} and possible future data on nucleon number
dependence on anomalous $J/\psi$ suppression. In this respect see
e.g. Ref.\cite{NPP}. The explanation of data on anomalous
 $J/\psi$ suppression by QGP will only be generally accepted provided
that the explanation will lead to experimentally verified predictions
concerning the onset as function of A,B and $E_T$. 
\item   Additional information on pre-resonance interaction with
nucleons, see e.g. Refs.\cite{BM,satz,KS95,KS2}
 (''inverse experiment''
and antiproton interactions in nuclei).
\item Making use of recent results of lattice calculations
\cite{Karsch}, which are becoming more and more accurate and lead to
the critical temperature for the phase transition of about $173$MeV.
This represents a constraint for models with $J/\psi$ suppression
by a gas of hadrons, since the energy density of the hadron gas higher
than $\epsilon_{QGP}(T_c)$ indicates the transition of the hadron gas 
into the QGP.
\end{itemize}

\medskip
{\bf Acknowledgements}\\
 One of the authors (J.P.) is
indebted to the CERN Theory Division for the hospitality 
extended to him. We would like to thank F. Karsch, C.Salgado and
B. Tom\'a\v{s}ik  
for comments and helpful discussions.

\end{document}